**PHYSICAL SCIENCES: Physics**

**Order parameter fluctuations at a buried quantum critical point**

Yejun Feng[1,2*], Jiyang Wang[2], R. Jaramillo[3], Jasper van Wezel[4], S. Haravifard[1,2], G. Srajer[1], Y. Liu[5, 6], Z.-A. Xu[6], P. B. Littlewood[7], and T. F. Rosenbaum[2*]

(1) The Advanced Photon Source, Argonne National Laboratory, Argonne, Illinois 60439, USA

(2) The James Franck Institute and Department of Physics, The University of Chicago, Chicago, Illinois 60637, USA

(3) School of Engineering and Applied Sciences, Harvard University, Cambridge, Massachusetts 02138, USA

(4) Materials Science Division, Argonne National Laboratory, Argonne, Illinois 60439, USA

(5) Department of Physics, Pennsylvania State University, University Park, Pennsylvania 16802, USA

(6) Department of Physics, Zhejiang University, Hangzhou 310027, P. R. China

(7) Cavendish Laboratory, University of Cambridge, Cambridge CB3 OHE, UK

Author contributions: Y.F., R.J. and T.F.R. designed the experiment. Y.L. and Z.-A.X. provided samples. Y.F., J.W., S.H. and G.S. performed the measurement. J.v.W. and P.B.L. provided theoretical input. Y.F., R.J., and T.F.R. analyzed the data, Y.F., R.J., J.v.W. and T.F.R. prepared the figures and the manuscript, and all authors commented.

The authors declare no competing financial interests.

* To whom correspondence should be addressed. Email: yejun@aps.anl.gov; t-rosenbaum@uchicago.edu

**Abstract:**

Quantum criticality is a central concept in condensed matter physics, but the direct observation of quantum critical fluctuations has remained elusive. Here we present an x-ray diffraction study of the charge density wave (CDW) in $2H$-NbSe$_2$ at high pressure and low temperature, where we observe a broad regime of order parameter fluctuations that are controlled by proximity to a quantum critical point. X-rays can track the CDW despite the fact that the quantum critical regime is shrouded inside a superconducting phase, and, in contrast to transport probes, allow direct measurement of the critical fluctuations of the charge order. Concurrent measurements of the crystal lattice point to a critical transition that is continuous in nature. Our results confirm the longstanding expectations of enhanced quantum fluctuations in low-dimensional systems, and may help to constrain theories of the quantum critical Fermi surface.

**Text:**

A continuous change of phase often involves critical fluctuations that destabilize one phase in favor of another. These fluctuations characterize the nature of the phase transition, but can be difficult to measure directly. This difficulty is especially acute in broad classes of materials with quantum phase transitions [1-2], from colossal magnetoresistance manganites [3] to heavy fermion and cuprate superconductors [4-5] to archetypal, metallic ferromagnets [6-7], where strong interactions can cut off the critical behavior via a structural instability, or competing ground states can shroud the quantum critical point.

Charge- and spin-density wave (CDW/SDW) systems have been shown to be good candidates for experimental studies of quantum critical behavior, where fluctuations disrupt electron pairing and restore the metallic Fermi surface [8-9]. In these systems the interaction strengths are weaker than in strongly correlated materials, reducing the likelihood of strong first-order transitions and allowing experimental access to the quantum critical point. Recent low temperature studies of the spin-density-wave transition in bulk, elemental Cr under pressure demonstrated a continuous quantum phase transition in an antiferromagnetic metal [10-11], but the quantum fluctuation regime deduced via transport measurements was very narrow. Stronger fluctuations over a broader range are expected in systems with lower electronic dimensionality. Moreover, quantum criticality in two dimensional layered systems with predilections for density wave distortions have received sustained interest due to the observation of density wave pairing in the high-$T_C$ superconductors [12-14]. Here we present a low-temperature and high-pressure synchrotron x-ray diffraction study of the two-dimensional charge-density-wave system $2H$-NbSe$_2$, where scattering from the incommensurate charge order is possible even deep within the coexisting superconducting ground state. Our results demonstrate a wide regime of spatial fluctuations of the charge density wave order parameter that are manifestly quantum in nature.

$2H$-NbSe$_2$ is a model two-dimensional, charge-density-wave system [15-21] in which negatively charged electrons and positively charged holes form a spatially periodic arrangement below the transition temperature $T_{CDW} = 33.5$ K [15-21]. NbSe$_2$ also superconducts below $T_{sc} = 7.2$ K, and the electronic bandstructure supports both the CDW and superconductivity with some competition for electronic states [16, 22-23]. The electron and hole states that condense into the CDW comprise only 1% of the total density of states at the Fermi surface [18-19, 21] and both the CDW order parameter amplitude and its static phase coherence length $\xi_s$ are unaffected when the superconductivity is suppressed in a high magnetic field [16], suggesting that the superconducting state has minimal effect on CDW pairing. However, long-wavelength CDW amplitude modes do affect the superconductivity by modulating the total density of states at the Fermi surface [17, 22-23]. The established pressure-temperature (*P-T*) phase diagram [17, 24] is summarized in Fig. 2a. NbSe$_2$ remains superconducting at pressures as high as 20 GPa [24]. The anticipated pressure-driven CDW quantum phase transition below 5 GPa is therefore buried inside the superconducting phase and is inaccessible to electrical transport measurements [17]. This leaves x-ray diffraction as the technique best suited for probing the quantum critical CDW.

Scattering experiments with x-rays or neutrons are sensitive both to the ordered state and to its fluctuations. Elastic scattering measures the static form factor $S(q, \omega=0)$ which includes contributions from long range order in the form of Bragg peaks, as well as diffuse scattering from thermally (and quantum mechanically) populated dynamical modes (Fig. 1a). The population of critical fluctuations can be obtained from diffuse scattering, which can be used to determine both the fluctuation spectrum (*e.g.* mode softening) and the fluctuation correlation length $\xi_F$ approaching a critical point (Fig. 1b) [25, 26]. In the case of weak charge order at high pressure it is technically difficult to measure the diffuse scattering intensity, mostly due to the enhanced background scattering from the pressure environment. Similarly, inelastic scattering, which measures the full dynamic form factor $S(q, \omega)$ and can directly probe $\xi_F$, remains technically unfeasible in extreme sample environments. Fortunately, information about fluctuations can still be obtained from the Bragg diffraction peaks coming from the static order parameter. Within a mean field approach, the static coherence length $\xi_s$ of the ordered state is infinite, resulting in an arbitrarily sharp Bragg diffraction peak at the CDW wavevector $\boldsymbol{Q}$ at all points on the phase diagram within the ordered phase. This is seen in experiments as a longitudinally instrument resolution-limited peak at momentum transfer $\boldsymbol{k} = \boldsymbol{Q}$. In the presence of fluctuations the ordered state is affected by collective dynamical modes with $\boldsymbol{k} \neq \boldsymbol{Q}$ which reduce $\xi_s$ as the critical point is approached (Fig. 1c). The suppression of the static coherence length $\xi_s$ is thus a consequence of dynamic fluctuations, and always accompanies the divergence of the dynamical fluctuation correlation length $\xi_F$. The critical broadening of the elastic Bragg peak is most easily observed in incommensurate charge- or spin-density wave systems. At ambient pressure and approaching the thermal phase boundaries from below, diffraction line broadening and the corresponding reduction of $\xi_s$ have been observed in systems such as one-dimensional NbSe$_3$ [27], two-dimensional NbSe$_2$ [16], and three-dimensional Cr [28]. The measurements reported here are analogous to these previous experiments, with the important difference being that by varying a non-thermal parameter we can approach a *quantum* critical point from within the ordered state.

**Results**

We have explored the evolution of both the CDW and the crystal lattice at temperatures $T = 3.5$ and 12 K over a wide pressure range, $0 \leq P \leq 8.6$ GPa (arrows in Fig. 2a). At ambient pressure, the periodicity of the CDW is incommensurate with the crystal lattice. The CDW wave vector $\boldsymbol{Q}$ is approximately equal to $0.98\boldsymbol{a}$*/3 and, unlike other low-dimensional CDW systems [15, 29], does not experience a lock-in transition to the commensurate $\boldsymbol{a}$*/3 state. Here $\boldsymbol{a}$* is the (1, 0, 0) reciprocal lattice vector, and the CDW incommensurability $\delta$ is defined by $\boldsymbol{Q} = (1-\delta)(1/3, 0, 0)$. Due to the hexagonal crystal symmetry, CDW states in transition metal dichalcogenides typically possess a three-fold degeneracy. In the related compound $2H$-TaSe$_2$, these three CDW orientations coexist within the same spatial volume to form a "triple-$Q$" state, as observed by neutron and x-ray diffraction [15, 29]. Diffraction patterns for a triple-$Q$ state have not been reported to date in $2H$-NbSe$_2$ [15, 16, 29], and our own result shows the CDW state in NbSe$_2$ to be of the single-$Q$ type with a spatially separated, three-fold domain structure.

The hexagonal lattice constants $a$ and $c$ evolve continuously with pressure up to at least 8.6 GPa, with no sign within our measurement sensitivity of a first order structural transition across the CDW phase boundary. We identify the critical pressure for the CDW quantum phase transition with a discontinuity in the slope of the c-axis lattice constant at $P_c = 4.6$ GPa (Fig. 2b). No discontinuity in slope occurs for the basal-plane lattice constant $a$, which is well characterized by a single-parameter Birch equation of state with modulus $82.1 \pm 0.7$ GPa. The $c$-axis compressibility $\beta_c = -\mathrm{d}ln(c)/\mathrm{d}P$ is reduced by a factor of two across the quantum phase transition, going from 0.97 %/GPa in the CDW phase to 0.48 %/GPa after crossing the phase boundary. This latter value is comparable to the compressibility of the basal plane, $\beta_a = -\mathrm{d}ln(a)/\mathrm{d}P = 0.36$ %/GPa, suggesting that the quantum phase transition may be accompanied by an increase in the effective dimensionality of the electronic structure. Alternatively, the difference in $c$-axis compressibility across the CDW phase boundary can be related to the inter-layer interaction between the CDW order parameters on adjacent planes. The observed increase in $c$-axis compressibility in the ordered phase would then be attributed to the ability of the CDWs to adjust their phase so that positively charged peaks in one plane sit atop negatively charged troughs in the plane below, thereby minimizing the Coulomb interaction between the two dimensional density waves.

In addition to measuring lattice constants, high-resolution x-ray diffraction is sensitive to minute changes in lattice symmetry. In Fig. 2c we plot the widths of the (1, 1, 0), (0, 1, 0), and (0, 1, 1) diffraction peaks as a function of pressure; these three diffraction orders carry information on the equality between the basal plane lattice constants $a$ and $b$, the angle between the $a$- and $b$-axes, and whether the $c$-axis remains orthogonal to the basal plane. All three line widths remain close to the resolution limit with a lattice correlation length at least 1500 Å at all pressures, are consistently well modeled by the pseudo-Voigt line shape (Fig. 3b), and, most notably, show no signs of broadening across the CDW phase boundary. Hence the lattice remains hexagonal at all measured pressures and does not undergo a phase transition at the CDW quantum critical point. Additional measurements of the (0, 2, 0), (0, 2, 1), and (0, 2, 2) diffraction peaks yield the same result. The apparent insensitivity of the crystal lattice to the electronic phase may be attributed to weak CDW-phonon coupling in $NbSe_2$; the phonon dispersion spectrum measured by inelastic scattering shows broad and extremely weak phonon softening near both $a*/3$ and $Q$ [15, 30].

We traced the evolution of the CDW from ambient pressure to $P \sim P_c$ at both $T = 3.5$ and 12 K. At base $T$, we could no longer discern CDW diffraction peaks for $P > 4$ GPa at our sensitivity level of $10^{-7}$ times the (1, 1, 0) lattice peak intensity. We plot in Fig. 3 both the CDW and the corresponding lattice peaks at five pressures approaching the quantum phase transition. The lattice peaks remain nearly resolution limited at all $P$. As discussed above, this suggests that the crystal lattice is insensitive to the encroaching CDW transition. However, the CDW diffraction peaks measured from the same samples under identical conditions are substantially broadened as the system is tuned towards the critical point, corresponding to a decrease of the static coherence length $\xi_s$ in real space.

The CDW wave vector $Q$ remains incommensurate throughout the $P$-$T$ phase diagram despite the proximity of the $\boldsymbol{a}$*/3 commensurate position (dashed line in Fig. 3a). The generic Ginzburg-Landau expression for the free energy of a CDW order parameter $|\psi|e^{i\varphi}$ contains two competing terms which determine the ordering wave vector. The first term is proportional to $|\psi|^2|\nabla\varphi-\mathbf{q}|^2$ and describes the energetic cost for repopulating electrons and holes in the case $Q$ does not coincide with the optimal Fermi surface nesting vector $q$ as determined by the band structure [31]. This is balanced by the Umklapp term proportional to $|\psi|^n\cos(n\varphi)$, which describes the energy gained by having a commensurate Q = $\boldsymbol{a}$*/$n$ [31]. In many materials the interplay between these terms leads to a discontinuous incommensurate-commensurate phase transition [15, 31], which becomes more likely as the order parameter grows. In 2$H$-NbSe$_2$, the CDW-lattice coupling is weak and even in the low temperature limit at ambient pressure, where the order parameter is the strongest, the CDW avoids this so-called lock-in transition.

At $P$ = 4 GPa and $T$ = 3.5 K the static coherence length $\xi_s$ of the CDW order parameter has decreased to approximately 26 Å (Fig. 3a), equivalent to the size of eight unit cells or about three CDW wavelengths. It is necessary to ask whether this decrease in coherence length could result from effects other than the proliferation of CDW fluctuations. Quenched disorder could in principle cut off the critical behavior. We do in fact find that the CDW correlation length is limited by strong pinning disorder at ambient pressure. However, as $P_c(T\rightarrow 0)$ is approached from below the diffraction line shape indicates a change in the nature of spatial fluctuations and the system effectively approaches the clean limit. This development is illustrated in Fig. 4. The longitudinal CDW line shape at $P = 0$ (top) fits well to a Lorentzian-squared functional form with a $(\Delta q)^{-4}$ tail, indicating that the 600 Å static coherence length $\xi_s$ of the CDW is constrained by quenched disorder and strong pinning Lee-Rice domains [32]. Under the application of pressure, the line shape (bottom) changes to a Lorentzian with a characteristic $(\Delta q)^{-2}$ tail; the broad line width at $P = 2.08$ GPa corresponds to a static coherence length $\xi_s$ of 110 Å. The Lorentzian line shape is not connected to strong pinning by quenched disorder, but rather arises from spatial fluctuations that are controlled by proximity to the critical point [27, 32]. Our counting statistics permit us to follow the distinction between Lorentzian and Lorentzian-squared forms up to $P = 2.6$ GPa.

We can also address the possible effects of the CDW domain structure on the measured line shapes. As described above, the three-fold rotational symmetry in the basal plane results in three CDW domain types. Since our measurement cannot directly measure domain size, one must consider the possibility that the reduced CDW coherence length at high pressure is the result of a proliferation of domain walls. Here our detailed line shape analysis presented in Fig. 4 also applies and could rule this possibility out. Any series of closely spaced and spatially abrupt disruptions of the CDW wavefront will produce a Lorentzian-squared line shape. At ambient pressure we ascribed this line shape to quenched disorder and the formation of Lee-Rice domains [32]. However, the same line shape would result from finite domain size. This analysis has long been applied to determine crystallite size in power diffraction [33]. In particular, it is well known that finite size broadening results in a line shape that is indistinguishably close to Lorentzian-squared, and that can be clearly distinguished from Lorentzian with a high resolution measurement. The line shapes presented in Fig. 4 plainly show that the static order parameter at high pressure is not limited by finite domain size, and instead is characterized by an exponential correlation length $\xi_s$, as is expected for a stochastic limiting process such as critical fluctuations.

Anisotropic stress in the pressure cell could influence $\xi_s$ through CDW-lattice coupling. The anisotropic strain in our samples is evidenced by the increase in the crystal mosaic width during pressurization (Methods). However, even at the highest measurement pressure of 8.6 GPa the lattice Bragg peaks remain resolution-limited longitudinally, corresponding to a lower bound of 1500 Å for the lattice correlation length (Fig. 2c). The difference of (at least) two orders of magnitude between the lattice correlation length and the CDW correlation length (26 Å) establishes that the anisotropic stress in our pressure cell is not responsible for the observed CDW line shapes.

**Discussion**

We now address the connection of the observed order parameter fluctuations to the CDW phase diagram and the quantum phase transition. In Fig. 5a we show data collected as a function of pressure at both low ($T$ = 3.5 K) and intermediate ($T$ = 12 K) temperatures. As illustrated in Fig. 3 and quantified in Fig. 5a, the divergence of the line width with pressure dominates any variation due to temperature; even though the curves in Fig. 5a differ by more than a factor of three in temperature, points at equal reduced pressure show only minimal thermal broadening. The divergence as the critical point is approached therefore cannot be ascribed to thermal fluctuations. Taking the same data, but plotting in Fig. 5b the inverse line width (*i.e.* the static coherence length $\xi_s$ of the CDW) as a function of reduced temperature, the same conclusion is obtained. Even though the measurements made at $T = 3.5$ K clearly are much farther from the thermal phase transition than the measurements made at $T = 12$ K, their coherence length at any given reduced temperature is much shorter. The fluctuations evidenced by the CDW line width cannot be caused by thermal fluctuations, and are instead controlled by proximity to the quantum critical point.

For any finite temperature, measurements sufficiently close to the phase boundary will show the effects of thermal fluctuations. In Fig. 5b we also show the evolution of the CDW coherence length as a function of temperature at ambient pressure. In this case the coherence length is limited by strong pinning for reduced temperatures $t = 1 - T/T_c > 0.2$. Closer to the phase boundary (for $t < 0.2$, or $T > 27$ K) thermal fluctuations control the coherence length and the line shape approaches a Lorentzian. In this context we note that our measurements at $T = 3.5$ K all correspond to $t > 0.6$ (Fig. 5b), thus reinforcing the conclusion that the thermal critical regime does not affect the line widths measured at high pressure and low temperature.

The onset of line width broadening at $P \sim 2$ GPa $\sim 0.4\ P_c$ suggests a broad regime of quantum fluctuations, reminiscent of the fluctuating quantum ground states observed near magnetic quantum critical points in other itinerant electron systems [6-7]. These fluctuations are likely connected to the electronic structure of NbSe$_2$ [18-21]. The small patches of Fermi surface that support the CDW are not well nested [21], and density functional theory predicts that the non-interacting susceptibility is characterized by a broad plateau rather than a narrow maximum at the observed $\boldsymbol{Q}$ [20]. Therefore the CDW spectrum may be generically soft, and it is reasonable to expect a wide regime of CDW fluctuations even far from the critical point. It is difficult experimentally to identify a boundary between an imperfectly nested ground state and the onset of quantum criticality. However, we point out that at the highest measured pressure of 4.0 GPa the CDW diffraction line corresponds to a broadening of about 3.6% in reciprocal space around the nesting vector Q, as determined by comparing the FWHM of the critical CDW line to the magnitude of the reciprocal lattice vector $\boldsymbol{a^*}$ (Fig. 3a). This is unlikely to be attributed to the imperfect nesting in the non-critical electronic spectrum of $NbSe_2$ [18-21], and is suggestive of the sizable critical fluctuations that are expected on the approach to the quantum critical point. Critical fluctuations over nearly all of reciprocal space were observed in the quantum critical heavy Fermion metal $CeCu_{1-x}Au_x$ [34], and were attributed in that case to local quantum criticality [35]. However, such a scenario is not expected to be applicable for $NbSe_2$ because it falls into the weak-coupling regime.

Strong fluctuations at the two-dimensional CDW quantum phase transition are a longstanding theoretical prediction [8, 9] that we here support experimentally. However, a full theory of two-dimensional coupled systems of electrons and damped collective modes at the CDW quantum phase transition remains to be formulated. The existing theory does predict an intermediate scaling regime of critical CDW fluctuations on the disordered side of the phase boundary [8], and the fluctuations that we observe for $2 < P < 4$ GPa suggest a similar scaling regime on the ordered side. For a generic incommensurate CDW, the scaling regime surrounding the quantum phase transition is predicted to be cut off by a discontinuous phase transition [8]. However, because the associated energy scale is proportional to the size of the incommensurability [8], the non-scaling regime may be expected to be very small in the case of NbSe$_2$. Our measured line widths should also help to constrain theories of collective mode softening on the ordered side of the quantum critical point.

Here we have shown that there exists a broad regime on the ordered side of *P-T* phase diagram of the two-dimensional CDW system 2*H*-NbSe$_2$ in which fluctuations at low temperatures limit the static coherence length of the CDW order parameter. These fluctuations are controlled by proximity to the quantum critical point, and are not thermal critical fluctuations due to the finite measurement temperature. Our data represent a rare direct observation of spatial fluctuations near a buried quantum critical point, as opposed to local or averaged probes in frequency. As the critical point is approached, the inverse static coherence length of the CDW diverges and the observed fluctuations become critical modes. Ongoing experimental efforts will help to resolve the universal character of this quantum critical point. Measurements of the diffuse elastic line width in the disordered phase could determine the critical exponent $\upsilon$ that controls the correlation length of fluctuations through $\xi_F \sim \left( P - P_c \right)^{-\upsilon}$ [36]. These measurements will depend on greatly improved counting statistics to allow high fidelity measurements of the diffuse elastic line shape on both sides of the phase boundary.

**Methods**

Each 50-μm thick NbSe$_2$ sample was blade-diced and loaded in the diamond anvil cell (DAC) in a methanol:ethanol (4:1 ratio) pressure medium with the basal plane parallel to the diamond culet. The crystal mosaic was 0.4$^o$ FWHM at ambient pressure and increased continuously to 2$^o$ FWHM at high pressure. A helium-controlled membrane was used to tune the pressure in the DAC [37]. The base temperature of 3.5 ± 0.2 K was achieved using a Gifford-McMahon cryocooler (Sumitomo RDK-205E) mounted on the sample stage of the x-ray diffractometer.

The pressure environment in this type of assembly has been studied extensively [37, 38]. The pressure anisotropy is estimated to be 0.01-0.02 GPa, while the pressure inhomogeneity is apparently random across the pressure chamber and depends on the sample size [37]. From Ref. [37] we estimate it is around ±0.02 GPa for our NbSe$_2$ samples (Fig. 2 inset) at $P_c$ = 4.6 GPa.

Experiments were performed at x-ray beamline 4-ID-D of the Advanced Photon Source, Argonne National Laboratory. Diffraction was in the vertical plane and in the transmission geometry to the *a-b* plane of NbSe$_2$. A double-bounce Si (1, 1, 1) monochromator selected 18.85-keV x-rays, chosen below the Nb K-edge to avoid this fluorescence excitation. The Se K-shell fluorescence was strongly absorbed by the diamond anvils, and also discriminated against by the acceptance window setting of the NaI scintillation detector. Two Pd-coated Kirkpatrick-Baez mirrors focused the x-ray beam and rejected higher harmonics. The crystal mosaic of our samples was larger than both the instrument resolution and the intrinsic transverse CDW widths [16, 28]. Thus our measured longitudinal CDW line shapes include an effective integration of the peak profile along both transverse directions. In spite of this complication, longitudinal scans of peaks such as (0, 2-*Q*, 0), (0, 1, 0) and (0, 2, 0) that are aligned along a single reciprocal lattice vector ensured a high precision measurement of the CDW wave vector *Q*. Given the sample mosaic under pressure, we were unable to measure CDW correlation lengths in both transverse directions, which are known at ambient *P* [16].

**References:**

[1] Hertz JA (1976) Quantum critical phenomena. *Phys Rev B* 14:1165-1184.

[2] Millis AJ (1993) Effect of a nonzero temperature on quantum critical points in itinerant fermion systems. *Phys Rev B* 48:7183-7196.

[3] Salamon MB, Jaime M (2001) The physics of manganites: Structure and transport. *Rev Mod Phys* **73**, 583-628.

[4] Park T, *et al.* (2006) Hidden magnetism and quantum criticality in the heavy fermion superconductor CeRhIn$_5$. *Nature* 440:65-68.

[5] Cooper RA, *et al.* (2009) Anomalous Criticality in the Electrical Resistivity of La$_{2-x}$Sr$_x$CuO$_4$. *Science* 323:603-607.

[6] Pfleiderer C, *et al.* (2007) Non-Fermi liquid metal without quantum criticality. *Science* 316:1871-1874.

[7] Uemura YJ, *et al.* (2007) Phase separation and suppression of critical dynamics at quantum phase transitions of MnSi and (Sr$_{1-x}$Ca$_x$)RuO$_3$. *Nature Phys* 3:29-35.

[8] Altshuler BL, Ioffe LB, Millis AJ (1995) Critical behavior of the $T = 0$ $2k_F$ density-wave phase transition in a two-dimensional Fermi liquid. *Phys Rev B* 52:5563-5572.

[9] Metlitski MA, Sachdev S (2010) Quantum phase transitions of metals in two spatial dimensions. II. Spin density wave order. *Phys Rev B* 82: 075128.

[10] Jaramillo R, *et al*. (2009) Breakdown of the Bardeen-Cooper-Schrieffer ground state at a quantum phase transition. *Nature* 459:405-409.

[11] Jaramillo R, Feng Y, Wang J, Rosenbaum TF (2010) Signatures of quantum criticality in pure Cr at high pressure. *Proc Natl Acad Sci USA* 107:13631-13635.

[12] Tranquada JM, *et al*. (1995) Evidence for stripe correlations of spins and holes in copper oxide superconductors. *Nature* 375:561–563.

[13] Hayden SM, Mook HA, Dai P, Perring TG, Dogan F (2004) The structure of the high-energy spin excitations in a high-transition-temperature superconductor. *Nature* 429:531-534.

[14] de la Cruz, C. *et al*. (2008) Magnetic order close to superconductivity in the iron-based layered $LaO_{1-x}F_xFeAs$ systems. *Nature* 453:899–902.

[15] Moncton DE, Axe JD, DiSalvo FJ (1977) Neutron scattering study of the charge-density wave transition in $2H$-$TaSe_2$ and $2H$-$NbSe_2$. *Phys Rev B* 16:801-819.

[16] Du C-H, *et al*. (2000) X-ray scattering studies of $2H$-NbSe$_2$, a superconductor and charge density wave material, under high external magnetic fields. *J Phys Condens Matter* 12:5361-5370.

[17] Berthier C, Molinié P, Jérome D (1976) Evidence for a connection between charge density waves and the pressure enhancement of superconductivity in 2H-NbSe$_2$. *Solid State. Comm*. 18:1393-1395.

[18] Straub Th, *et al*. (1999) Charge-Density-Wave Mechanism in $2H$-NbSe$_2$: Photoemission Results. *Phys Rev Lett* 82: 4504-4507.

[19] Rossnagel K, *et al*. (2001) Fermi surface of $2H$-NbSe$_2$ and its implications on the charge-density-wave mechanism. *Phys Rev B* 64:235119.

[20] Johannes, M.D., Mazin, I.I. & Howells, C.A. (2006) Fermi-surface nesting and the origin of the charge-density wave in NbSe$_2$. *Phys Rev B* 73:205102.

[21] Borisenko SV, *et al*. (2009) Two Energy Gaps and Fermi-Surface "Arcs" in NbSe$_2$. *Phys Rev Lett* 102:166402.

[22] Littlewood PB, Varma CM (1981) Gauge-Invariant Theory of the Dynamical Interaction of Charge Density Waves and Superconductivity. *Phys Rev Lett* 47:811.

[23] Littlewood PB, Varma CM (1982) Amplitude collective modes in superconductors and their coupling to charge-density waves. *Phys Rev B* 26:4883.

[24] Suderow H, Tissen VG, Brison JP, Martínez JL, Vieira S (2005) Pressure Induced Effects on the Fermi Surface of Superconducting 2H-NbSe$_2$. *Phys Rev Lett* 95:117006.

[25] Collins MF (1989) *Magnetic Critical Scattering*. (Oxford University Press, New York).

[26] Holt M, Sutton M, Zschack P, Hong H, Chiang T-C, (2007) Dynamic Fluctuations and Static Speckle in Critical X-Ray Scattering from SrTiO$_3$. *Phys. Rev. Lett.* 98:065501.

[27] DiCarlo D, Thorne RE, Sweetland E, Sutton M, Brock JD (1994) Charge-density-wave structure in NbSe$_3$. *Phys Rev B* 50:8288-8296.

[28] Hill JP, Helgesen G, Gibbs D (1995) X-ray-scattering study of charge- and spin-density waves in chromium. *Phys Rev B* 51:10336.

[29] Fleming RM, Moncton DE, McWhan DB, DiSalvo FJ (1980) Broken Hexagonal Symmetry in the Incommensurate Charge-Density Wave Structure of 2*H*-TaSe$_2$. *Phys Rev Lett* 45:576-579.

[30] Schmalzl K, Strauch D, Hiess A, Berger H (2008) Temperature dependent phonon dispersion in 2H-NbSe$_2$ investigated using inelastic neutron scattering. *J Phys Condens Matter* 20:104240.

[31] McMillan WL (1975) Landau theory of charge-density waves in transition-metal dichalcogenides. *Phys Rev B* 12:1187-1196.

[32] Ravy S, *et al*. (2006) Disorder effects on the charge-density waves structure in V- and W-doped blue bronzes: Friedel oscillations and charge-density wave pinning. *Phys Rev B* 74:174102.

[33] Jones FW (1938) The measurement of particle size by the x-ray method. *P. Roy. Soc. A* 166:16-43.

[34] Schröder A, *et al.* (2000) Onset of antiferromagnetism in heavy-fermion metals. *Nature* 407:351-355.

[35] Si Q, Rabello S, Ingersent K, Smith JL (2001) Locally critical quantum phase transitions in strongly correlated metals. *Nature* 413:804-808.

[36] Moudden AH, Axe JD, Monceau P, Levy F (1990) $q_1$ charge-density wave in NbSe$_3$. *Phys. Rev. Lett.* 65:223.

[37] Feng Y, Jaramillo R, Wang J, Ren Y, Rosenbaum TF (2010) High-pressure techniques for condensed matter physics at low temperature. *Rev. Sci. Instrum*. 81:041301.

[38] Jaramillo R, *et al.* (2008) Chromium at high pressures: weak coupling and strong fluctuations in an itinerant antiferromagnet. *Phys Rev B* 77:184418.

**Acknowledgments**

We thank X. Lin for help on sample growth. The work at the University of Chicago was supported by NSF Grant No. DMR-0907025. Use of the Advanced Photon Source and the work at the Material Science Division of Argonne National Laboratory were supported by the U.S. Department of Energy (DOE) Basic Energy Sciences under Contract No. NE-AC02-06CH11357. The work at Zhejiang University was supported by NSF of China.

**Figure Captions**

**Fig 1.** Correlation lengths of both the ordered state and its fluctuations as measured by elastic scattering. (A) Schematics of a typical peak profile across the transition temperature $T_c$. The scattering profile is composed of both the Bragg diffraction of the static order below $T_c$ (blue), and the diffuse scattering of dynamical fluctuations at all temperatures (red). The relative intensity of the two components varies with system and temperature, and also depends on the choice of diffraction technique and instrument resolution settings. Assuming a Lorentzian profile for both components, the measured full peak widths at half maximum (FWHM) are related to correlation lengths $\xi$ by FWHM=$2/\xi$ [27]. (B) The correlation length $\xi_F$ of the dynamic fluctuations diverges with the same critical exponent on both sides of $T_c$. (C) The static coherence length $\xi_s$ of the ordered state is always infinite up to $T_c$ on the mean-field level. However, taking into account fluctuation effects causes $\xi_s$ to become finite near the thermal phase transition. The corresponding broadening of the Bragg peak becomes measurable in a high-resolution diffraction setup.

**Fig. 2.** Phase diagram and crystal lattice of NbSe$_2$ across the pressure driven CDW quantum phase transition. (A) $P$-$T$ phase diagram of NbSe$_2$. The CDW transition temperature $T_{CDW}$ (open squares, Ref: [17]; solid square, our data) drops sharply under pressure while the superconducting transition $T_{SC}$ (circles, Ref: [17, 24]) is nearly constant. Arrows at $T$ = 3.5 and 12 K mark our experimental trajectories. The CDW phase boundary inside of the superconducting phase is found to be $P_c$ = 4.6 GPa. (Inset) Micrograph shows a typical pressure chamber assembly. In addition to the NbSe$_2$ sample, a ruby ball and annealed polycrystalline silver were included as manometers for use at room-$T$ and low-$T$, respectively. (B) Lattice constants of hexagonal 2$H$-NbSe$_2$ as a function of pressure at $T$ = 3.5 K, plotted against the lattice constant of silver measured *in situ* [37]. $a$ and $c$ are refined from six measured diffraction orders: (1, 1, 0), (0, 1, 0), (0, 2, 0), (0, 1, 1), (0, 2, 2), and (0, 2, 1). The two vertical axes are scaled to represent the same range of relative compression ($\approx$7.5%) for both $a$ and $c$. Above the CDW the two compression curves run nearly parallel, suggesting a change of electronic dimensionality from 2D to 3D. (C) Measured full width at half maximum (FWHM) of longitudinal scans of (1, 1, 0), (0, 1, 0), and (0, 1, 1) diffraction orders as a function of pressure; the corresponding dashed lines mark the instrument resolution.

**Fig. 3.** Pressure evolution of the lattice and CDW diffraction line shapes. Longitudinal scans of the CDW and lattice Bragg peaks at $T$ = 3.5 K at 5 pressures approaching the quantum phase transition under identical conditions. The CDW line width broadens quickly as $P$ nears $P_c$, while the Bragg peaks remain resolution limited. Data are vertically displaced with each peak height normalized to unity. Solid lines are fits to the data. The vertical dashed line indicates the commensurate position (0, 5/3, 0). Vertical scale bars next to the CDW data represent intensity

ratios $I(0, 2\text{-}Q, 0)/I(0, 2, 0)$, where $I(0, 2\text{-}Q, 0)$ and $I(0, 2, 0)$ are the peak intensities of the corresponding CDW and lattice peaks at the same pressure. The $(0, 2, 0)$ Bragg reflection is one of the weakest in $NbSe_2$, and is about 2500 times less intense than $(1, 1, 0)$. The lattice line shape is best characterized by a pseudo-Voigt function, indicating comparable contributions from the sample lattice and the instrument resolution. At high pressure the CDW line shape becomes much broader than the instrument resolution (note that the two $x$-axis scales differ by a factor of 7.5) and is best fit with a symmetrical Lorentzian function on top of a linear background. We observe no evidence for asymmetry in the CDW line shape. The Lorentzian line shape provides direct evidence of CDW fluctuations approaching the quantum phase transition. Vertical error bars correspond to $1\sigma$ uncertainty in the counting statistics.

**Fig. 4.** Change of the CDW line shape from ambient to high pressure. Both longitudinal scans were fit with two types of functional form, Lorentzian (black) and Lorentzian squared (red) plus a constant background. The data at ambient pressure shows a second harmonic peak of the CDW at $(0, 1/3+2\delta, 0)$, which was observed in $2H$-$TaSe_2$ [15] but never reported for $2H$-$NbSe_2$; this observation testifies to the high quality of the crystals used in this study. The Lorentzian-squared line shape at ambient pressure indicates a CDW correlation length dominated by disorder [29]. By contrast, the Lorentzian functional form at $P = 2.08$ GPa is typical of fluctuation broadening and demonstrates that our crystals are in the effectively clean limit at the approach to the quantum critical point. A linear sloped background was removed from the raw scan at 2.08 GPa (cf. Fig. 2a) to highlight the tail region. Vertical error bars correspond to $1\sigma$ uncertainty in the counting statistics.

**Fig. 5.** CDW fluctuations and the *P-T* phase diagram of 2*H*-NbSe$_2$. (A) FWHM of CDW longitudinal scans for both *T* = 3.5 (blue) and 12 K (purple) as a function of pressure. Data taken from different samples use different symbols. Projected onto a color map of reduced pressure 1-*P*/*P*$_c$(*T*) in the *P-T* phase diagram, these two trajectories (arrows) at different *T* clearly experience a similar variation in reduced pressure. (B) The static coherence length $\xi_s$ of the CDW as a function of reduced temperature, 1-*T*/*T*$_c$(*P*); data were recorded as a function of pressure at *T* = 3.5 K (blue) and *T* = 12 K (purple), and as a function of temperature at *P* = 0 (red). Arrows point in the directions of increasing *T* and *P*. Projecting onto a color map of 1-*T*/*T*$_c$(*P*) in the *P-T* phase diagram makes clear that the *T* = 12 K trajectory experiences a much wider variation in reduced temperature than does the *T* = 3.5 K trajectory. Solid lines are guides to the eye.

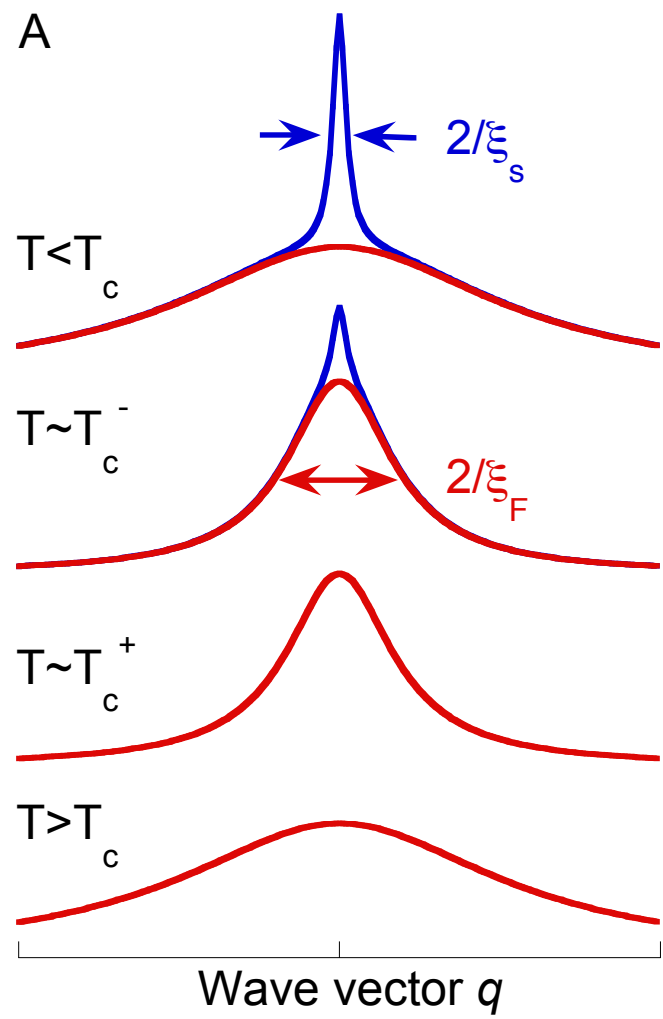

A

$T < T_c$

$2/\xi_s$

$T \sim T_c^-$

$2/\xi_F$

$T \sim T_c^+$

$T > T_c$

Wave vector $q$

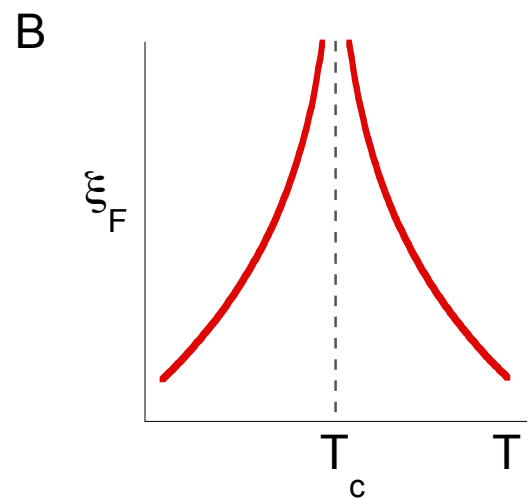

B

$\xi_F$

$T_c$    $T$

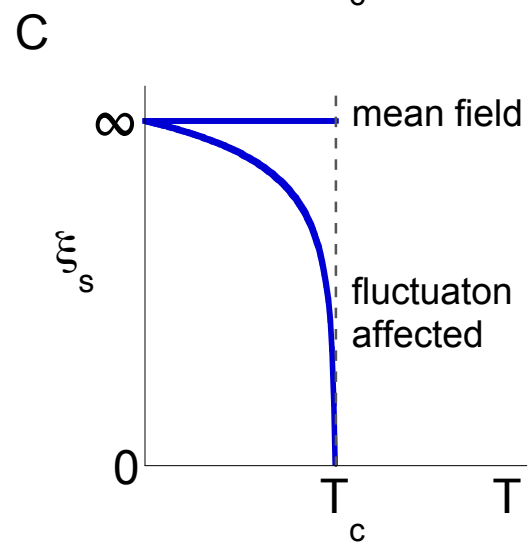

C

$\xi_s$

$\infty$

mean field

fluctuaton affected

$T_c$    $T$

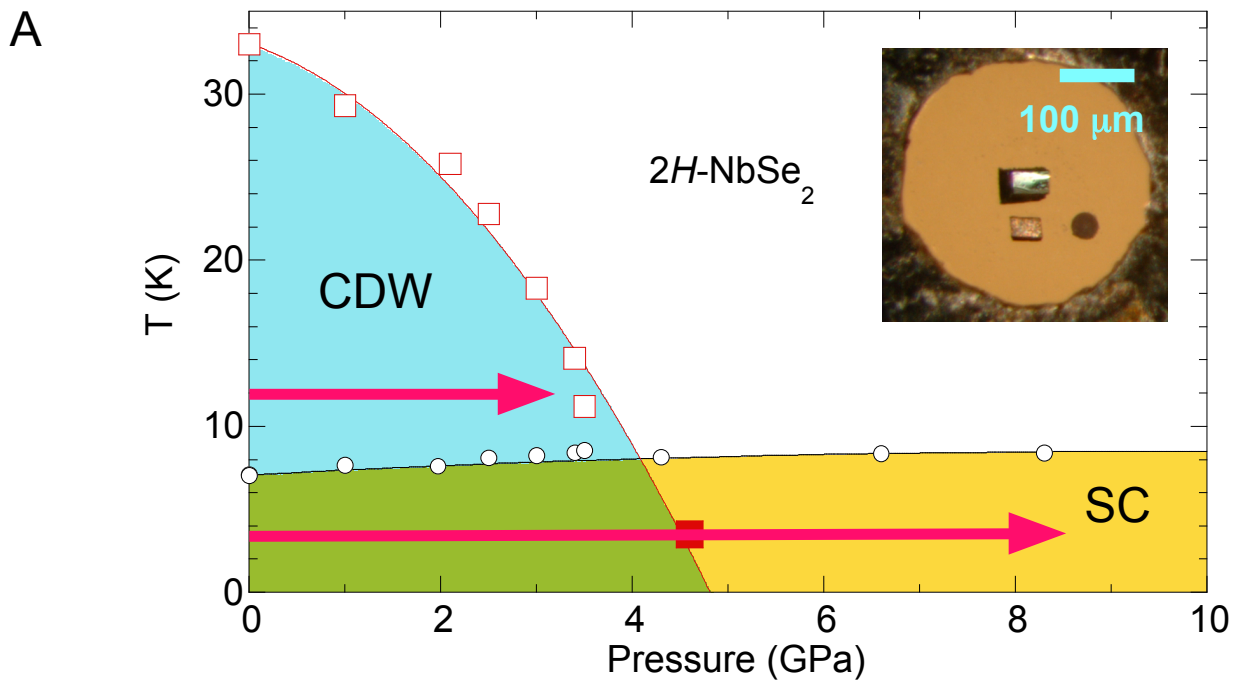

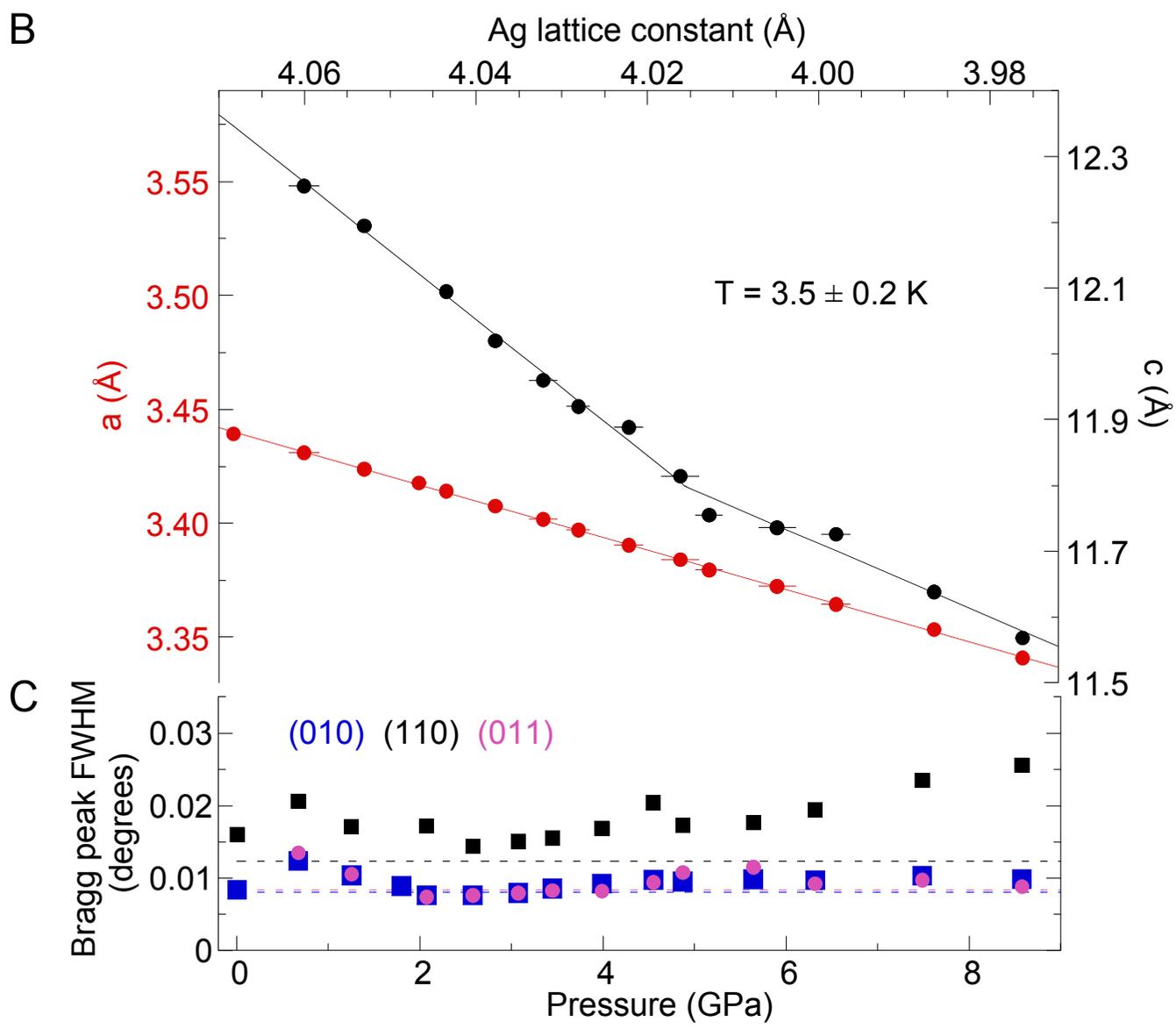

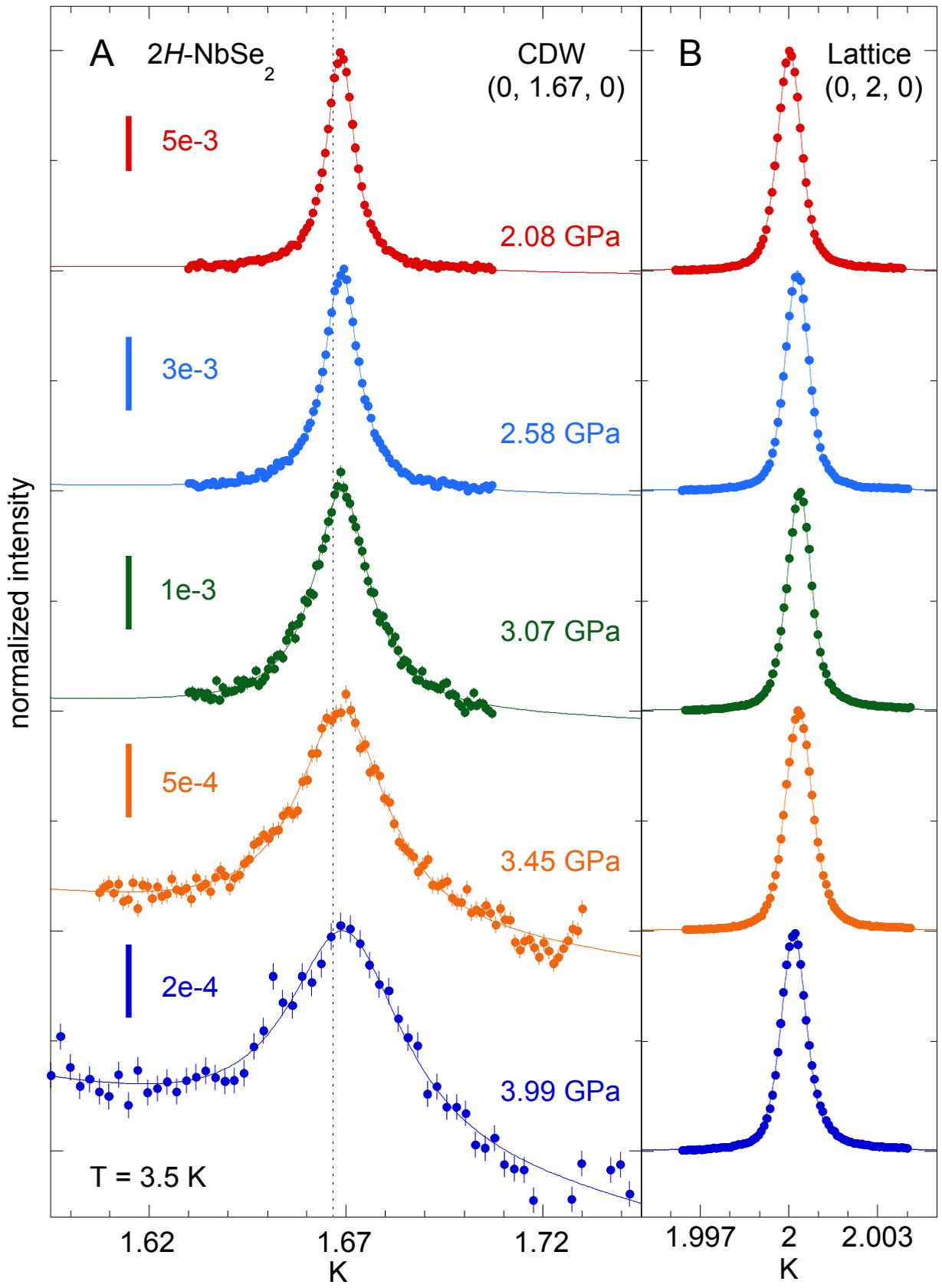

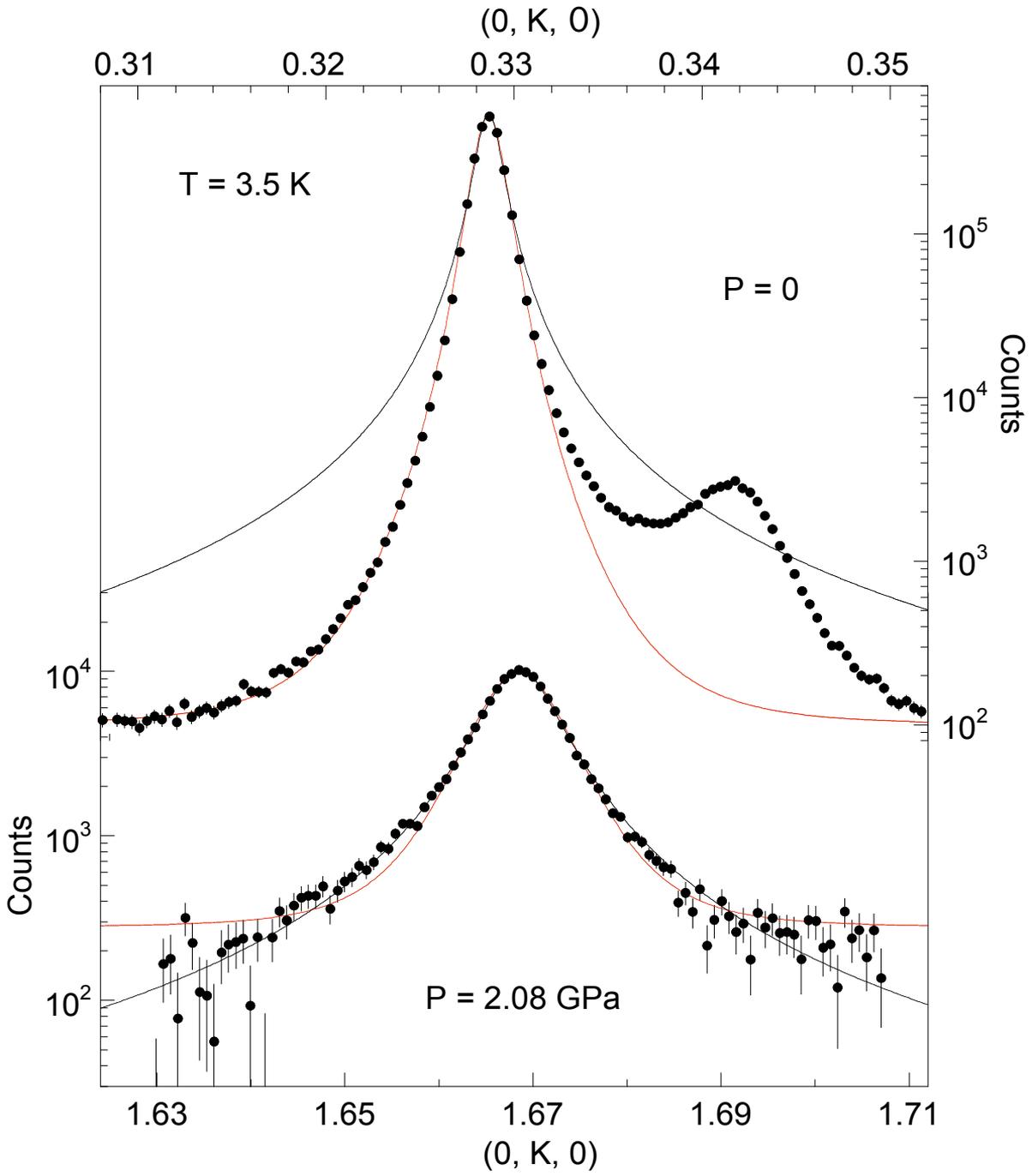

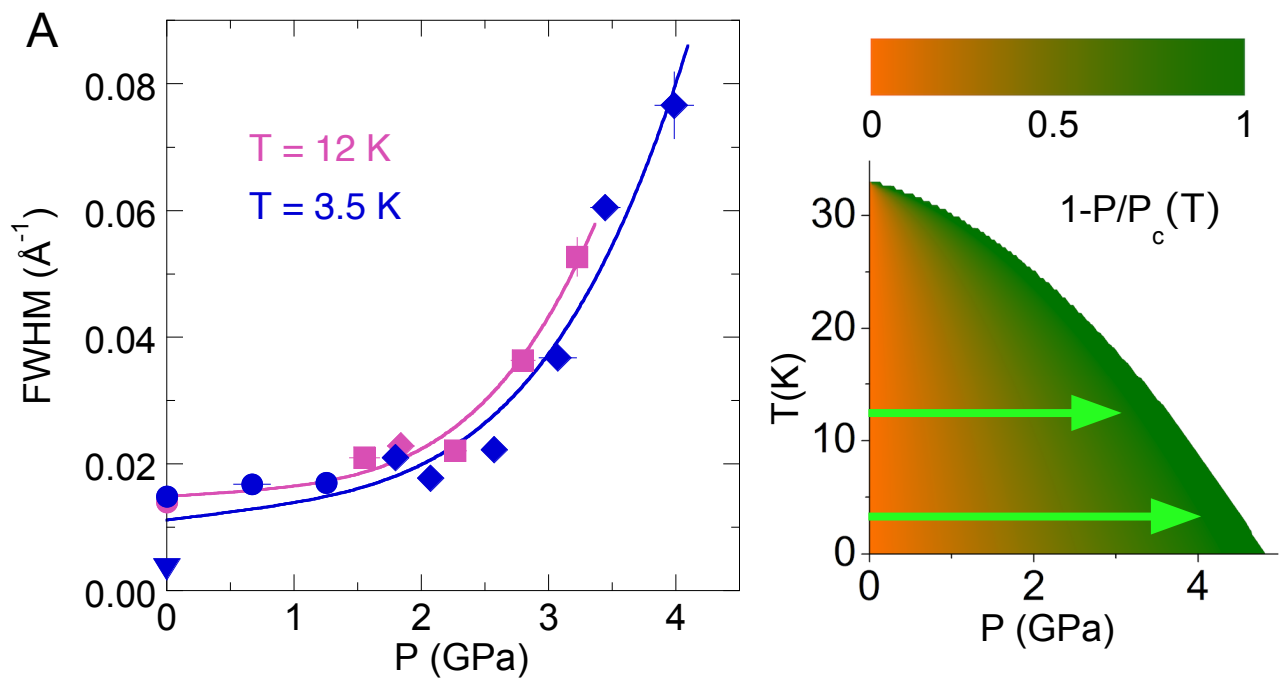

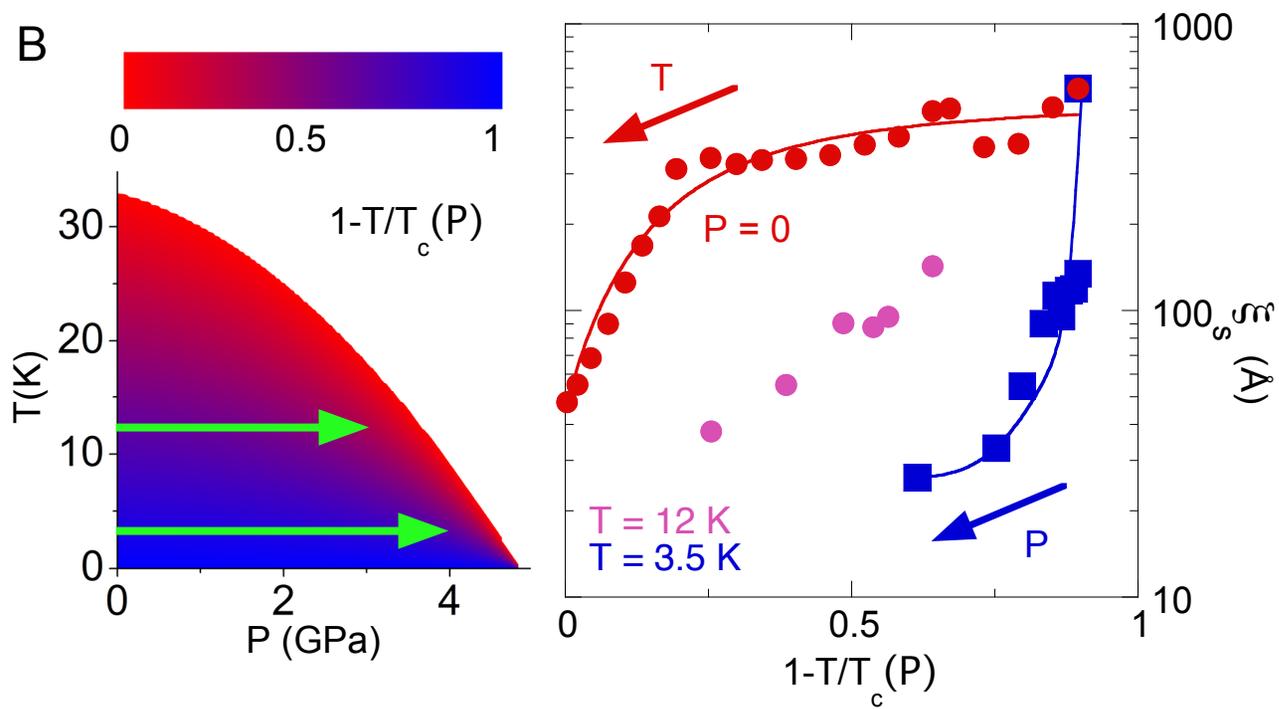